\documentclass[12pt]{article}  
\usepackage{amsmath}
\usepackage{amsfonts,amssymb}
\usepackage{psfig}

\def\1{{\mathchoice {1\mskip-4mu\mathrm l}
{1\mskip-4mu\mathrm l} {1\mskip-4.5mu\mathrm l}
{1\mskip-5mu\mathrm l}}}

\newtheorem{theorem}{Theorem}[section]

\title{Chaotic temperature dependence at zero temperature.}

\author{
{\normalsize A.C.D. van Enter  }\\
{\normalsize Center for Theoretical Physics,} \\
{\normalsize University of Groningen},\\
{\normalsize Nijenborgh 4, }\\
{\normalsize 9747AG,Groningen},\\
{\normalsize  the Netherlands}\\
\\[3 mm]
{\normalsize W.M. Ruszel}\\
{\normalsize Center for Theoretical Physics,} \\
{\normalsize University of Groningen},\\
{\normalsize Nijenborgh 4, }\\
{\normalsize 9747AG,Groningen},\\
{\normalsize the Netherlands}}

\begin{document}
\maketitle

\newpage
\begin{abstract}
We present a class of examples of nearest-neighbour,
bounded-spin models, in which the
low-temperature Gibbs measures do not converge as the temperature is lowered
to zero, in any dimension.
\end{abstract}

\section{Introduction}

In most examples we know in statistical mechanics, ground states measures,
or at least a subset of the set of ground states measures, can be obtained
as proper limits of low-temperature states (Gibbs measures). In fact, the same
tends to be true at positive temperatures: changing the
temperature leads to a limit of Gibbs measures (in the weak
$*$-topology) approaching some other Gibbs measure at the new
temperature. It is not unususal  that a limit of pure (extremal) Gibbs measures
is non-pure, as often happens for low-dimensional ground states, and
as also at positive temperatures is known for example in the
``Thouless effect'' \cite{T,ACCN}. Also examples are known where there
exists an infinite sequence of first-order phase transitions at low
temperatures, but still there is the possibility of convergence to at least
some ground state as the temperature approaches zero, see e.g.
\cite{ES,PS}. In this note we present an example where there is {\em no} low-temperature limit of {\em any} sequence of Gibbs
measures for a prescribed sequence of temperatures. Indeed,
lowering the temperature will mean that one oscillates between
Gibbs measures having either a ferromagnetic or antiferromagnetic
character. Such a ``messy'' \cite{rue} behaviour seems an example
of the ``chaotic temperature dependence'', which has been proposed
in the spin-glass literature, see e.g.  \cite{NSc,NSs}. (We use here
the terminology ``chaotic'' for non-convergence, in the same spirit
as it was  introduced in the notions of ``chaotic size dependence'' and
``chaotic time dependence'' by Newman and Stein.)
Our example employs bounded spins with a bounded interaction. The
spins take continuous values, and we don't know if similar
behaviour can occur for discrete spins. Also, our interaction is
not continuous, but this is not an essential feature, and the example could
be made continuous. Although the notion of chaotic temperature dependence
was introduced in the spin-glass literature, our interactions do
not contain disorder, but are deterministic.

\section{Notation and Result}

We will say that Chaotic Time Dependence (CTD) holds, if no ground state
measure can be obtained as a proper limit of finite-temperature Gibbs
measures \cite{G} by lowering the temperatures.
(This does allow convergence on certain subsequences of temperatures, as indeed
necessarily happens due to compactness arguments).

We consider  for simplicity 2-component spins (the generalization to
N-component spins is immediate), which live on the
unit circle, and which are located on a d-dimensional lattice
$\mathbb{Z}^d$, with the following (formal) Hamiltonian:

\begin{equation}
H = - \sum_{<i,j> \in \mathbb{Z}^d} U(\theta_i - \theta_j)
\label{Hamiltonian}
\end{equation}

The potential $U$ is a sum of two ``Seuss'' potentials having the form
of wells-in-wells (hats-in-hats) \cite{ES,Seuss}, one
ferromagnetic and the other one  antiferromagnetic:

\begin{equation}
U(x) =  \sum_{n \geq 1} \Bigl ( \frac{1}{2^{2n}} +
\frac{1}{2^{2n+1}} \Bigr ) \1_{\varepsilon_{2n}}(x) + \sum_{n \geq
1} \Bigl ( \frac{1}{2^{2n+1}} + \frac{1}{2^{2n+2}} \Bigr )
\1_{\varepsilon_{2n+1}}(\pi + x) +
\frac{1}{4}\1_{\varepsilon_1}(\pi + x) \label{potential}
\end{equation}
with $\varepsilon_n = \varepsilon^{3^{n}}$ for $\varepsilon$ small
enough. The steps (wells) in the two sums have asymmetric widths
($\varepsilon_{2n}$ resp. $\varepsilon_{2n+1}$) and heights but
the maximum of both terms (the two sums of the depths  of the wells) is
$\frac{1}{2}$. The first sum contains the ferromagnetic terms, the
second sum antiferromagnetic ones.

\medskip

The construction is such that, precisely as in \cite{ES}, there
is, in $d$ at least 2, an infinite sequence of first-order transitions to ever
deeper and narrower wells. However, now the successive wells
alternate between being ferromagnetic and antiferromagnetic.

Lowering the temperature $T$ for this potential means that the
typical bond configurations jump between successive wells centered at
$0$ and $\pi$, i.e. between a ferromagnetic resp.
antiferromagnetic state.

\medskip

We can therefore construct a sequence of temperatures $T_n$ (or equivalently
 inverse temperatures $\beta_n$), such that for any choice of
$\alpha$ the limit $\mu_{\beta_n}^{\alpha}$
 does not exist as $T_n$ goes to $0$, that is $\beta_n$ goes to $\infty$.

\begin{theorem}
Let $U$ be defined as in $\eqref{potential}$. For $N$-vector
models in any dimension and $N \geq 2$, chaotic temperature-dependence
 (CTD) occurs at $T=0$.
\end{theorem}

For $d$ at least $2$, the proof is an immediate corollary of
\cite{ES}. Indeed, take a sequence of temperatures such that the
typical bonds for ({\em any}) choice of  Gibbs measure are oscillating between
ferromagnetic and antiferromagnetic, corresponding to the
even-numbered  and odd-numbered  wells in \cite{ES}. Then there
are (at least) two limit points, a ferromagnetic and an
antiferromagnetic one. In $d$ at most $2$, presumably there will
be only two, (the rotation invariant mixtures of the pure
configurations), in $d$ at least $3$, magnetized and N\'eel
ordered ground state configurations are possible subsequence
limits of various low-temperature Gibbs measures.

However, the phenomenon does not need the existence of these phase
transitions and indeed occurs already in one dimension.

For the proof in $d=1$ we present an elementary argument. We look
at the bond distributions at different bonds. Each bond
independently takes values on the circle, describing the
difference in spin values between neighboring sites.
The bonds are thus i.i.d. on the circle $[0,2\pi)$, so it suffices
to look at the probability $P_n(\beta)$ of one bond being in
precisely the $n$-th well at  inverse temperature $\beta$ and to
show that for  appropriately  chosen $\beta_n$ and $n$ this
probability is larger than one half. Then either the probability
that the system is in the ferromagnetic state (most bonds
ferromagnetic) or the probability that it is in the
antiferromagnetic state (most bonds antiferromagnetic) is larger
than one half and the system keeps oscillating between those two
if $\beta$ increases. \newline

The probability that the bond is in the $n$-th well and not
in the $n+2$-th well (by construction it is not in the $n+1$-th
well) is

\begin{equation*}
P_n(\beta) = \frac{1}{Z_{\beta}}(\varepsilon_n -
\varepsilon_{n+2})\exp \biggl( - \frac{3 \beta}{2^{n+1}}\biggr)
\end{equation*}
where $Z_{\beta}$ is a normalizing constant depending on the
inverse temperature $\beta$. The probability $P_n(\beta)$, considered as
a function of $n$, is maximal iff its logarithm
( equals minus the n-th well's ``free energy'')

\begin{equation*}
-f_{\beta}(n):= -\frac{3 \beta}{2^{n+1}} + \ln(\varepsilon_n -
\varepsilon_{n+2})
\end{equation*}
is maximal. As usual, maximal probability corresponds to minimal free energy.
The derivative of $f$ with respect to $n$ (now taken
as a continuous variable) is equal to

\begin{equation*}
f^{\prime}_{\beta}(n)= c_1 \frac{\beta}{2^n} - c_2(\varepsilon)3^n
\end{equation*}
with $c_i$ strictly positive constants, $c_2$ depending on
$\varepsilon$. Then it follows immediately that $f$ has its
minimal value at $n_{max}$ which satisfies
\begin{equation}
6^{n_{max}} = \frac{c_1}{c_2(\varepsilon)} \beta . \label{max}
\end{equation}

It is clear that increasing $\beta$ means increasing
$n_{max}$ to get the most probable position of the bond. Choosing
a sequence of inverse temperatures $\beta_{n} = O(6^{n})$ and
$\varepsilon$ appropriately (such that $n_{max}=n$ is an integer) we
get that the position of most probable well $n_{max}$ is alternating between
even and odd, i.e. the bond concentrates itself
alternatingly either at $0$ or $\pi$.

{\bf Remark:} If the minimum of $f$ is at a value between integers,
there can be a two-fold degeneracy, with wells $n$ and $n+1$ being
equally probable.

It remains to prove that $P_{n_{max}}(\beta_{n}) \geq \frac{1}{2} + \delta$
for some small number $\delta$, 
for all suffiently large $n$, which is equivalent to

\begin{equation*}
P_{n_{max}}(\beta_{n}) \geq \sum_{n \neq n_{max}} P_{n}(\beta_{n})+ \delta
\end{equation*}

or

\begin{equation*}
(\varepsilon_{n_{max}} - \varepsilon_{n_{max}+2})\exp \biggl( -
\frac{3 \beta_n}{2^{n+1}} \biggr) \geq \sum_{n \neq n_{max}}
(\varepsilon_{n} - \varepsilon_{n+2})\exp \biggl( - \frac{3
\beta_{n}}{2^{n+1}}\biggr) + \delta.
\end{equation*}

We prove the somewhat stronger result that the probability
distribition over the wells becomes more and more sharply peaked; indeed the 
probability of $n$ being the favourite well increases to one when $n$ and the 
corresponding inverse temperature $\beta_n$ increase to infinity.

For the estimate we first neglect that the wells lie inside each
other.
Once $\varepsilon$ is chosen small enough,
the mistake we make this way is sufficiently small that the inequality we will
derive remains true.

\bigskip

Assume thus that the wells are separate, and thus the probability of
being in the $n$-th well is

\begin{eqnarray*}
\tilde{P}_n(\beta) & = &\frac{1}{\tilde{Z}_{\beta}} \exp \biggl (
-
\frac{3 \beta}{2^{n+1}} - c_{\varepsilon}3^n \biggr ) \\
& = & \frac{1}{\tilde{Z}_{\beta}} \exp(- \tilde{f}_{\beta}(n))
\end{eqnarray*}

where $\tilde{Z}_{\beta}$ is again a normalizing constant and
$c_{\varepsilon} := -\log(\varepsilon)$.

Using the same argument as before we get that the most probable
position of the bond is at $n_{max}$ which satisfies

\begin{equation*}
\beta = c_{\varepsilon} \frac{2}{3} \frac{\log(3)}{\log(2)}
6^{n_{max}},
\end{equation*}

i.e. at the minimal point of the function $\tilde{f}_{\beta}(n)$.
(Here again we first consider the variable $n$ to be continuous
and then look at integer values for the maximal $n$, choosing the
sequence $\beta_{n}$ appropriately.) Then for $k \in \mathbb{N}$
note that

\begin{equation*}
\tilde{f}_{\beta_{n}} (n_{max}+k)
= 3^{n_{max}}c_{\varepsilon} (
2^{-k} a + 3^k )
\end{equation*}

resp.

\begin{equation*}
\tilde{f}_{\beta_{n}} (n_{max}-k)
= 3^{n_{max}}c_{\varepsilon} (
2^{k} a + 3^{-k})
\end{equation*}

with the abbreviation $a:= \frac{\log(3)}{\log(2)}$.

 Comparing the ratio $\frac{\tilde{P}_{n_{max} \pm k
}(\beta_n)}{\tilde{P}_{n_{max}}(\beta_{n})}$ we can show easily that


\begin{equation}
\frac{\tilde{P}_{n_{max} \pm k
}(\beta_{n})}{\tilde{P}_{n_{max}}(\beta_{n})} \leq
\exp(-c_{\beta_{n}} k) \label{estimate_ratio}.
\end{equation}.

When the constant $c_{\beta_{n}}$ is large enough this proves our
claim.

But the above statement  follows directly at low enough temperatures,
once we notice that

1) The function $\tilde{f}_{\beta} (n)$ is convex in $n$.

\noindent
and

2) The differences
$|\tilde{f}_{\beta_{n}} (n_{max}) \pm \tilde{f}_{\beta_{n}} (n_{max} \pm 1)|$
\noindent
diverge when  the sequence of inverse temperatures $\beta_{n}$ diverges;
in fact these differences diverge proportionally to  $3^{n}$.

\section{Final remarks and conclusion}
We have constructed a bounded-continuous-spin model with a
bounded interaction, and a sequence of temperatures
converging to zero, such that
the (any)  sequence of Gibbs measures at these temperatures does not converge
to a ground state. This seems to be the first example in which a form of
Chaotic Temperature Dependence has been proven.

In our example there are many more ground states than the ferromagnetic
and antiferromagnetic ones, however, by adding non-nearest neighbor terms,
one can suppress these if one wants.

Whether the phenomenon can also occur at positive temperatures, or for
discrete spins, at this point remain open questions.
However, it is known that for one-dimensional, sufficiently
short-range,   discrete-(finite-)spin interactions the Gibbs measures
do converge to a limit - ground state -  measure as the temperature 
decreases \cite{B,L}.

{\bf Acknowledgements:} We thank Christof K\"ulske for some helpful remarks on the manuscript.

\end{document}